\documentclass[%
twocolumn,superscriptaddress,
 amsmath,amssymb,
 aps,
prl,
]{revtex4-2}
\usepackage{mathrsfs,amsmath,amssymb,amsthm}
\usepackage{amssymb,fmtcount}
\usepackage{graphicx,epsfig,latexsym,overpic,amssymb,color}
\usepackage{threeparttable}
\preprint{\today}

\begin{document}

\title{
Survival Probabilities of Compound Superheavy Nuclei Towards Element 119
}
\author{Yu Qiang}
\affiliation{
State Key Laboratory of Nuclear Physics and Technology, School of Physics,
Peking University, Beijing 100871, China
}
\author{Xiang-Quan Deng}
\affiliation{
School of Physical Sciences, University of Chinese Academy of Sciences, Beijing 100049, China
}
\author{Yue Shi}
\affiliation{
Department of Physics, Harbin Institute of Technology, Harbin 150001, China
}
\author{C.Y. Qiao}
\affiliation{
Institute of Particle and Nuclear Physics, Henan Normal University, Xinxiang 453007, China
}
\author{Junchen Pei}\email{peij@pku.edu.cn}
\affiliation{
State Key Laboratory of Nuclear Physics and Technology, School of Physics,
Peking University, Beijing 100871, China
}
\affiliation{
Southern Center for Nuclear-Science Theory (SCNT), Institute of Modern Physics, Chinese Academy of Sciences, Huizhou 516000, China
}
\begin{abstract}
To synthesize superheavy element 119 is becoming highly concerned as several experimental projects in major laboratories are being pursued.
This work studied the survival probabilities of compound superheavy nuclei after multiple neutron emissions based on microscopic
energy dependent fission barriers, demonstrating a significant role of triaxial deformation in decreasing the first fission barriers in the heaviest region.
Together with the fusion cross sections by the dinuclear system model, the optimal energy and the residual cross section of $^{243}$Am($^{48}$Ca, 3$n$)$^{288}$Mc
are reproduced. Finally the cross sections and optimal beam energies of  $^{243}$Am+$^{54}$Cr and $^{249}$Bk+$^{50}$Ti reactions
are estimated, providing clues for the synthesis of new elements. 
\end{abstract}
\maketitle

\emph{Introduction.}---
To add new elements in the periodic table is a highly-concerned scientific question. 
To date, elements up to Z=118 have been synthesized~\cite{118}, 
mainly due to the adoption of the $^{48}$Ca projectile in fusion-evaporation reactions. 
However,  new elements in the 8th row of the periodic table 
can not be synthesized using the combination of
$^{48}$Ca and available targets, and it would be much more challenging. 
Experimental attempts to synthesize
Z = 119 and 120 elements were performed in laboratories using reactions such as $^{58}$Fe + $^{244}$Pu at JINR ~\cite{Fe58}, $^{51}$V + $^{248}$Cm
 at RIKEN~\cite{V51}, and $^{64}$Ni + $^{238}$U, $^{50}$Ti + $^{249}$Bk, $^{50}$Ti + $^{249}$Cf,
$^{54}$Cr + $^{248}$Cm at GSI ~\cite{gsi119,gsi119a,gsi120} , but no evidence of new elements was observed yet.
New facilities such as the SHE factory~\cite{Mc286} in Dubna and CAFE2~\cite{cafe2} in Lanzhou having very high beam intensities, also joined
the search for element 119. 
Due to limited choices of reaction systems, the reliable estimations of optimal
beam energies and cross sections are very desirable for such extremely difficult experiments. 

The fission and survival probabilities of highly excited compound superheavy nuclei are crucial
in designing the collision energies. 
Indeed, the energy dependencies of fission barriers are very different for different compound nuclei~\cite{pei2009}. 
For cold fusion compound nuclei, the fission barriers decrease rapidly with increasing excitation energies, and
the fission lifetimes would be smaller than that of hot fusion nuclei by 2 orders at high excitations~\cite{zhuy}. 
However, the energy dependence of fission barriers is conventionally taken into account by an empirical damping parameter in 
modelings of survival probabilities~\cite{itkis}. 
It should be very cautious to apply such empirical parameters to calculate cross sections of new reaction systems.
Furthermore,  the survival probability has an  ``arch" structure and is crucial in determining the optimal 
projectile energy together with the monotonically increasing fusion probability.

The first-chance survival probabilities of compound superheavy nuclei
have been studied previously based on microscopic energy dependent fission barriers~\cite{qiaocy},
including barrier heights and curvatures. 
This has not been applied in practical calculations of survival probabilities and cross sections after the
evaporation of multiple neutrons. 
The microscopic fission barriers~\cite{qiaocy} are much higher than that from macroscopic-microscopic (MM) models~\cite{moller} in the heaviest
region. To resolve this problem, the triaxial deformation should be considered, which plays a significant role to reduce the first barriers~\cite{caiqz,lvbn}. 
Note that the energy dependence of fission barriers can not be self-consistently described by MM models.
Besides, the reliable masses of superheavy nuclei are also very relevant for 
calculations of reaction Q values and survival probabilities~\cite{geng}, which are obtained recently by optimizing
Skyrme energy density functionals with $\alpha$-decay energies up to Z=118~\cite{guan}. 

In this Letter, we aim to study the optimal beam energies to produce element 119 based
on  microscopic survival probabilities, which are long-expected since the microscopic studies of 
fission barriers of compound superheavy nuclei more than a decade ago~\cite{pei2009}. 
The multiple neutron emissions are included so that energy dependent fission barriers of a cascade of superheavy isotopes  are calculated,
in which triaxial deformation and octuple deformation are included. 
Together with the fusion probabilities by the dinuclear system model~\cite{deng}, the
cross sections and optimal beam energies are predicted. 
For benchmark, the  $^{243}$Am($^{48}$Ca, 3$n$)$^{288}$Mc reaction~\cite{Mc286} is reasonably reproduced,
for which the latest experimental cross sections are available.

\emph{The Method.}---
The synthesis process of superheavy nuclei
can be described as the capture-fusion-evaporation reaction.
In this procedure the final residue cross section $\sigma_{ER,xn}$ is written as~\cite{deng}
\begin{equation} 
\begin{array}{ll}
\sigma_{{\rm ER},xn}(E_{\rm cm}) = & \sum_J \sigma_{\rm cap}(E_{\rm cm}, J)P_{\rm CN}(E_{\rm cm}, J) \vspace{5pt} \\
& \times W_{{\rm sur},xn}(E_{\rm cm}, J) \\
\end{array}
\end{equation}
where $E_{\rm cm}$ is the incident energy in the center-of-mass frame
and $J$ is the relative angular momentum.
$\sigma_{ER,xn}$ depends on the capture cross-section $\sigma_{\rm cap}$, the probability $P_{\rm CN}$ to form a compound nucleus, and the survival
probability $W_{{\rm sur},xn}$ of compound nuclei after evaporation of multiple neutrons.

The energy dependent fission barriers are calculated by the microscopic self-consistent
finite-temperature Hartree-Fock+BCS (FT-BCS) approach~\cite{goodman}, in which
the quantum shell effects and pairing correlations gradually disappear
as excitation energies increases. 
 The finite-temperature Hartree-Fock-Bogoliubov approach is computationally very costly,
 thus FT-BCS is adopted~\cite{zhuy}.
 The FT-BCS equation is solved by using the HFODD code~\cite{hfodd}, including the
 triaxial deformation and octupole deformation (see calculation details in Ref.~\cite{yshi}). 
 The Skyrme force UNEDF1~\cite{unedf1} is adopted for nuclear interaction, which
 can  reasonably describe fission barrier heights. 
 Beside, the density dependent mixed-type pairing interaction~\cite{mix-pair} is adopted. 
 In principle, the uncertainties of fission barriers can be obtained by Bayesian analysis~\cite{bayesian},
 but it is too expensive for three-dimensional calculations with energy dependence.

 To describe the experimental fission probabilities,
 the energy dependent fission barriers and level density parameters
 have to be employed~\cite{qiaocy}.
 The level density parameter $a_0$ at the ground state shape is usually taken as $A$/12, where
 $A$ is the nuclear mass number. Note that the level density parameter of $A$/12 has been widely adopted
 in modelings of synthesis of superheavy nuclei~\cite{zhang,deng,fengzq,xia}, and the consequences of  more complicated formulism of level density~\cite{zubov,Ignatyuk}
 involving additional parameters is deserved to be studied in the future.
 To estimate the uncertainties of level densities in this work, 
 results with  $A$/11,  $A$/12, and  $A$/13 are presented for comparison. 
  The level density parameter $a_0$
 is also used in calculations of neutron evaporation width.
  Beside, the level density
 parameter $a_f/a_0$ is taken as an adjustable parameter to
 reproduce experiments as $a_0$ varies.
 Previously our studies demonstrate that
 the main difference between the Bohr-Wheeler model
 and the imaginary free energy method (ImF) is because 
 the different level densities between the ground state and the saddle point~\cite{qiaocy}.
 The back-shifted Fermi-gas model is used to calculate the level density~\cite{fengzq}, including the dependence of angular momentum. 
 The fission probabilities are obtained using the hybrid model~\cite{qiaocy}, which is the combination of the Bohr-Wheeler model and
 the ImF method.

 
 The fusion cross sections are calculated with 
 the dinuclear system model (DNS).
The capture process is described as a
penetration through the Coulomb barrier between the nuclei. 
 By considering the evolution of dynamical
deformations of two colliding nuclei, the DNS with a dynamical potential energy surface
 was used to describe the complete
fusion probability by competing with quasifission~\cite{deng}.
Note that this process is modeled by a diffusion of the
DNS in the mass asymmetry degree of freedom, which is different from  the fission process in the nuclear deformation degree of freedom.
 The DNS model
has been widely applied in descriptions of fusion cross sections
in synthesis of superheavy nuclei~\cite{deng,fengzq,zhang,lijx,Nasirov,kayumov,fanli,adamian}. 
The masses of superheavy nuclei are taken from our latest microscopic calculations
to calculate reaction $Q$ values and neutron separation energies~\cite{guan}. 
The details of DNS calculations are given in a previous work~\cite{deng}.

\begin{figure}[ht]
\centering
\includegraphics[width=0.49\textwidth]{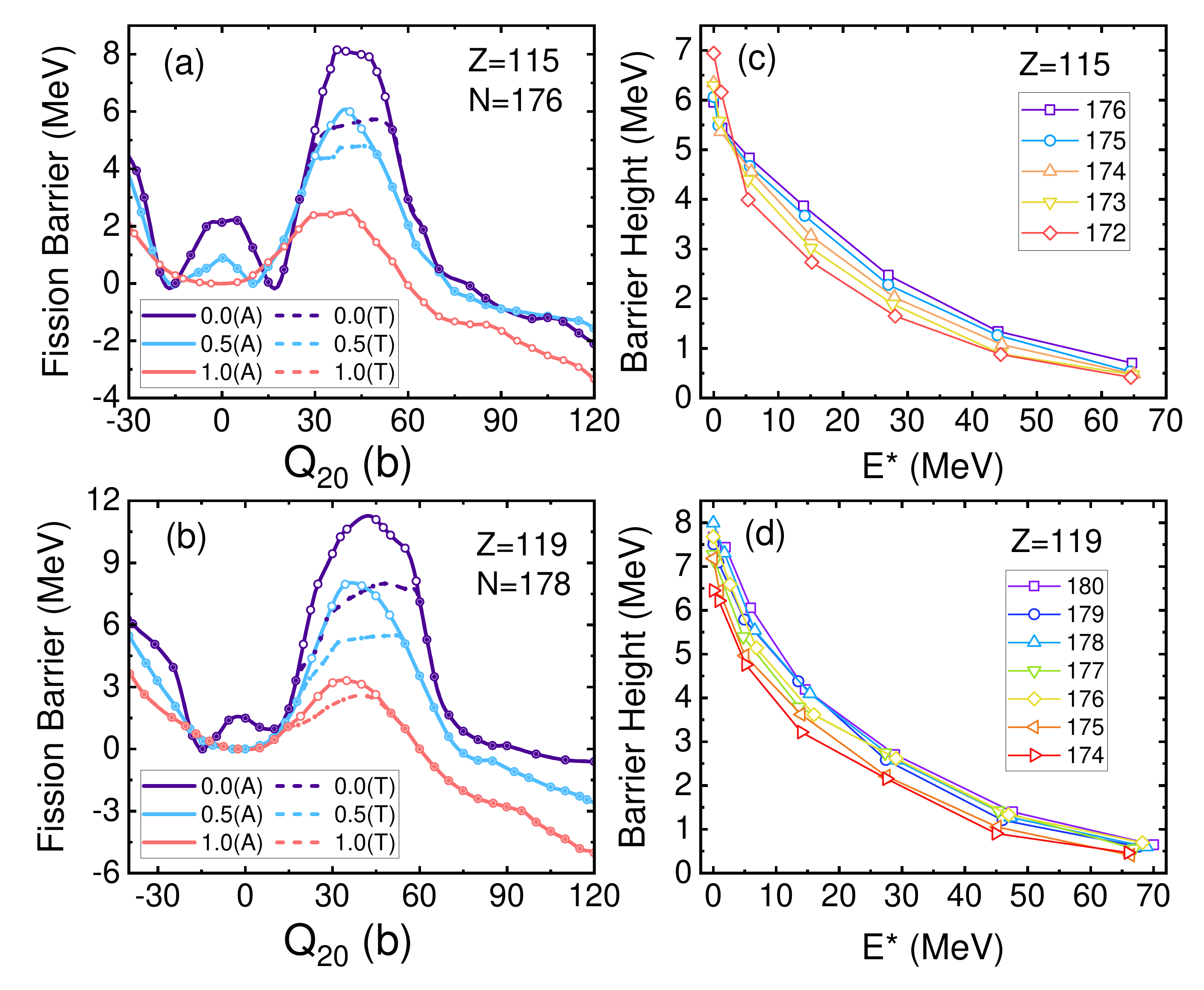}
\caption{
Calculated fission barriers of Z=115 (Mc) and Z=119 compound nuclei.
In subfigures (a) and (b), the fission barriers with triaxial deformation (T) of $^{291}$Mc and $^{297}$119 at different temperatures are shown, compared
with axial-symmetric fission barriers (A), which are displayed as a function of quardpole deformation $Q_{20}$.
In subfigures (c) and (d), the barrier heights of Z=115  and Z=119 isotopes are shown as a function of excitation energies E$^{*}$. 
                           \label{Fig2}
}
\end{figure}

\emph{The Results.}---
Firstly, the energy dependent fission barriers are calculated, as shown in Fig.\ref{Fig2}.
It is known that the cross sections are sensitive to the fission barriers, and
the optimal beam energies can vary by a few MeV~\cite{denisov}. 
The fission barriers of $^{291}$Mc and $^{297}$119 are shown in Fig.\ref{Fig2}(a, b), corresponding
to $^{243}$Am+$^{48}$Ca and $^{243}$Am+$^{54}$Cr reactions.
The second barriers disappear in both cases due to the octupole deformation. 
It can be seen that for $^{291}$Mc, the fission barrier height is 5.72 MeV with triaxial deformation, while
it is about 8.16 MeV with axial symmetry. The significant role of triaxial deformation in reducing the first barrier
has been discussed in transuranium nuclei, based on multidimensionally-constrained mean-field models~\cite{lvbn,yshi}.  The MM model calculations also indicate
that the triaxiality effect is more significant  towards 
neutron-rich superheavy nuclei at N$\geqslant$176~\cite{caiqz}. 
Indeed, we see that  triaxiality effect is more significant in $^{297}$119
and plays a role even at high excitations. 
The reduced role of triaxial deformation and octupole deformation at high extcitations has also been discussed previously~\cite{pei2009,sheikh}. 

The systematic results of energy dependent fission barrier heights of Z=115 and Z=119 nuclei are shown in Fig.\ref{Fig2}(c,d).
In the FRLDM calculations~\cite{moller}, the fission barrier heights of $^{291}$Mc and $^{297}$119 are 9.6 MeV and 7.94 MeV, respectively.
In another MM calculation, the fission barriers are around 6 MeV in this region~\cite{mm-barrier}. 
For Z=115, the heights of fission barriers are reasonable compared to MM calculations in Ref.~\cite{mm-barrier}. 
It can be seen that barrier heights generally become higher as the neutron number increases towards the N=184 shell. 
The fission barrier of $^{287}$Mc is higher than that of $^{291}$Mc due to reduced triaxiallity effect,
and becomes lower than that of $^{291}$Mc at high excitations. 
For Z=119 isotopes, however, 
the resulting microscopic fission barrier heights are higher than MM results, but are similar to FRLDM results. 
The fission barriers of Z=119 nuclei given by 
 the Weizs\"{a}cker-Skyrme (WS4) mass model are lower than 6 MeV~\cite{wangning}. 
Note that fission barriers obtained by UNEDF1 force are slightly lower than that by SkM* ~\cite{skm} force in this region. 
It was shown that different microscopic calculations predict that fission barriers around Z=119 are higher than that of Z=115~\cite{tub},
in contrary to MM calculations.
On the other hand, the fission barriers of Z=119 nuclei decreases more rapidly with increasing energies compared to 
that of Z=115 nuclei. 
At $E^{*}$$\thicksim$30 MeV, the barrier heights of Z=119 nuclei become close to that of Z=115 nuclei. 
Thus the higher fission barriers of Z=119 nuclei at zero temperature are not a problem for calculations of survival probabilities at high excitations.

\begin{figure}[ht]
\centering
\includegraphics[width=0.48\textwidth]{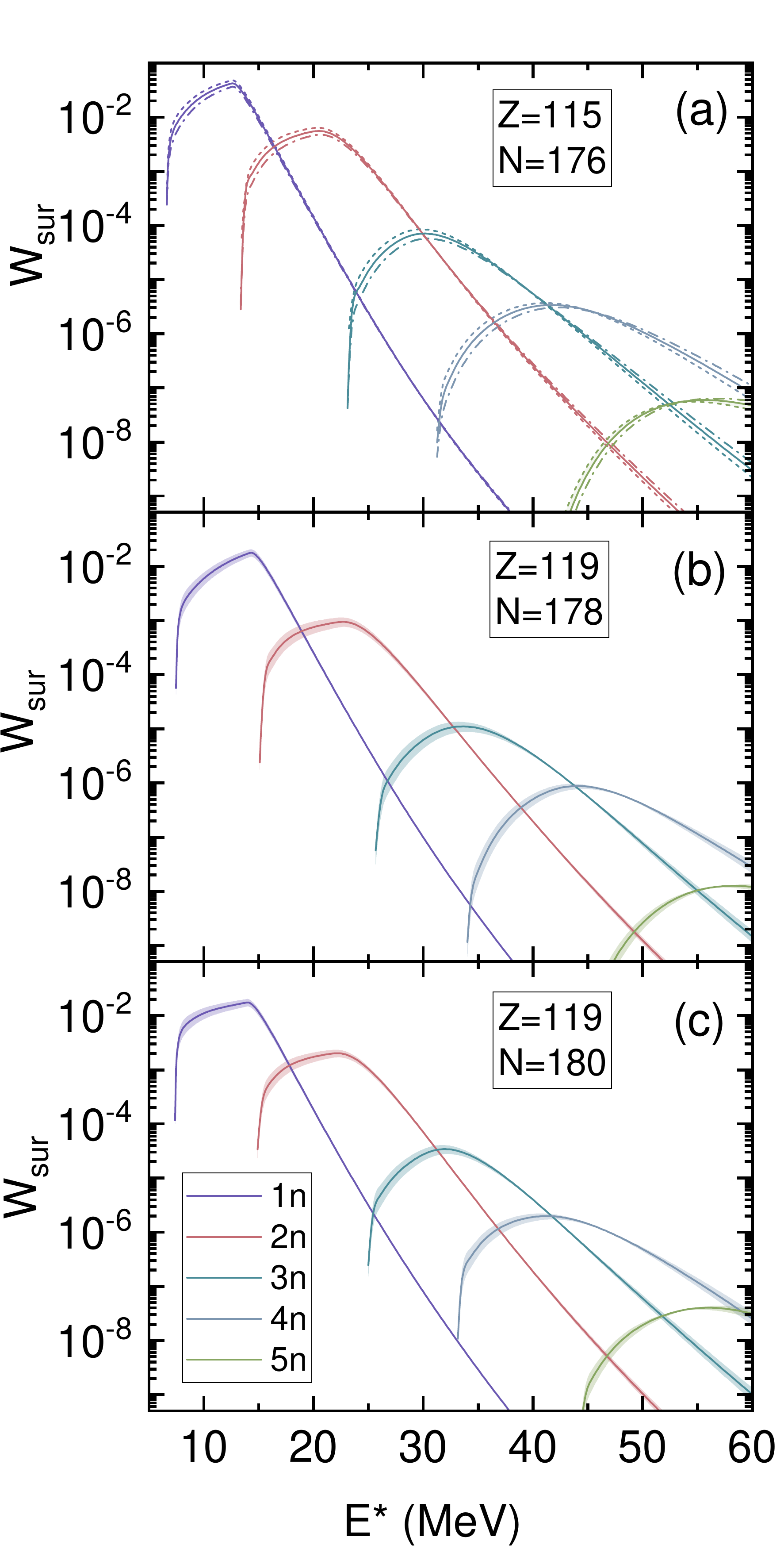}
\caption{
Calculated survival probabilities of compound nuclei $^{291}$Mc and $^{299, 297}$119
based on microscopic energy dependent fission barriers, after the evaporation of multiple neutrons. 
The uncertainties due to different level density parameters are shown.
                           \label{Fig3}
}
\end{figure}

\begin{figure}[ht]
\centering
\includegraphics[width=0.48\textwidth]{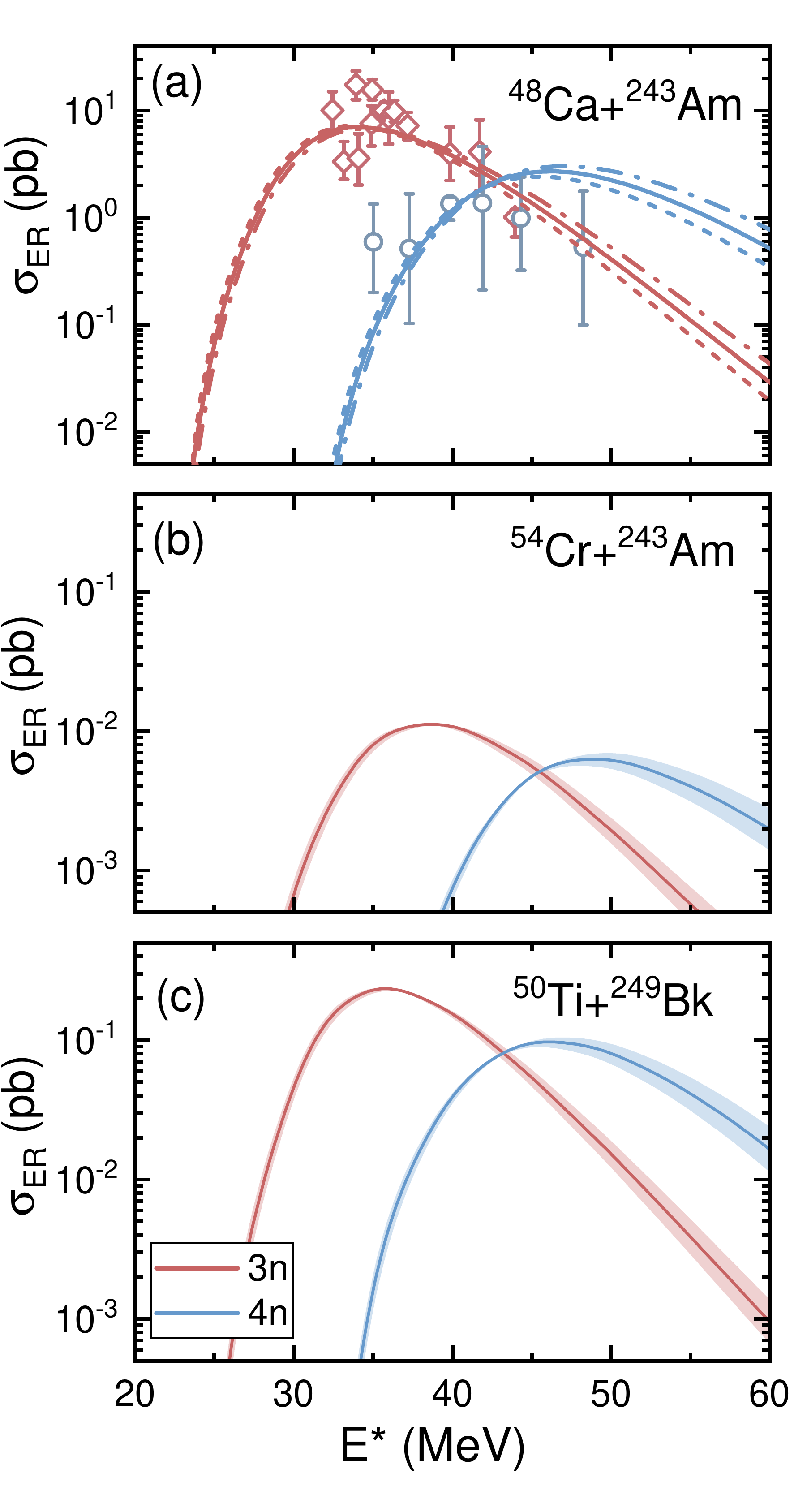}
\caption{
Calculated residual cross sections of  $^{243}$Am+$^{48}$Ca, $^{243}$Am+$^{54}$Cr and $^{249}$Bk+$^{50}$Ti reactions 
after the evaporation of multiple neutrons.  
The experimental data in the $^{243}$Am+$^{48}$Ca reaction are taken from \cite{Mc286}.
The shadow shows results due to different level density parameters. 
                           \label{Fig4}
}
\end{figure}

The survival probabilities of compound superheavy nuclei after multiple neutron emissions as a function of excitation energies
 are displayed in
Fig.\ref{Fig3}. 
The results with different level density parameters $a_0$=$A$/11, $A$/12 and $A$/13 are shown to estimate the uncertainties.
Note that $a_f$/$a_0$ are adjusted correspondingly to reproduce the residual cross sections of the $^{243}$Am+$^{48}$Ca reaction at 3$n$ channel.
 $a_f$/$a_0$ are taken as 1.07, 1.085, 1.1 for $a_0$=$A$/11, $A$/12 and $A$/13, respectively. 
 Then the effect due to different level density parameters is not significant. 
The neutron separation energy $S_n$ of $^{291}$Mc  is 6.77 MeV, which is larger than the barrier height of 5.72 MeV, then the 
first chance survival probability is very small. The first chance survival probability could be much larger
if the compound nucleus has an odd-number of neutrons. 
For the 3$n$ evaporation channel, the largest survival probability is around $E^{*}$$\sim$30 MeV, which is slightly smaller than 
the optimal excitation energy $\sim$34 MeV in experimental residual cross sections\cite{Mc286}, since the  monotonically increasing fusion probability would
suppress the low-energy contribution and
shift the optimal point to slightly higher energies. 
Therefore the energy of the largest survival probability provides a lower limit, irrespective of the capture and fusion process. 
The survival probabilities of $^{297}$119 are smaller than that of $^{299}$119, mainly due to its
lower fission barriers and larger $S_n$. 
For $^{291}$Mc, its survival probabilities are larger than that of $^{299, 297}$119 at 3$n$ channel due to its smaller $S_n$,
although its fission barriers are lower.

In self-consistent calculations of energy dependent fission barriers, the curvatures $\omega_0$ around the ground state are also energy dependent~\cite{zhuy}.
At high excitations, the survival probabilities can be enhanced significantly by multiplying the factor $\omega_0/T$ ($T$ denotes temperature)~\cite{qiaocy}, which is
included in the ImF method but not in the standard Bohr-Wheeler method. 
In this work, the factor $\omega_0/T$ is included in calculations of survival probabilities, where $\omega_0$ is 
 much decreased towards high excitations. 
To some extent, the factor $\omega_0/T$ compensates for the increasing dissipation effect at high excitations~\cite{qiangy}. 
This is supported by the observation of $^{243}$Am($^{48}$Ca, 5$n$)$^{286}$Mc at $E^{*}\sim$ 50 MeV~\cite{Mc286}.

Finally the residual cross sections of  $^{243}$Am+$^{48}$Ca, $^{243}$Am+$^{54}$Cr and $^{249}$Bk+$^{50}$Ti reactions are calculated
, as shown in Fig.\ref{Fig4}.
The updated cross sections of $^{243}$Am+$^{48}$Ca at 3$n$ channel are reproduced
by adopting the $a_f/a_0$=1.085. Such a factor $a_f/a_0$ is also used in calculations of Z=119 nuclei. 
The shadow region shows the uncertainties due to different level density parameters. 
The optimal excitation energy is around 33 MeV in the 3$n$ channel, which is very close to the experimental value around 34 MeV~\cite{Mc286}.
The cross section at 4$n$ channel is also reasonably reproduced. 
The estimated large cross section of 2$n$ channel is very large, however, 
the experimental cross section is small. This is also shown in Ref.~\cite{deng} with different fission barriers. 
Actually it is often  that calculations below the Bass barrier~\cite{bass} are too optimistic, as demonstrated by different reactions~\cite{oganessian24}.
Here the Bass barrier is around $E^{*}$=32 MeV.
Therefore we are focusing on 3n and 4n channels only, which are the most promising reactions.  

For the $^{249}$Bk+$^{50}$Ti reaction, the estimated cross sections at 3$n$ channel
is about 0.22 pb at $E^{*}$$\sim$35 MeV. 
In the experiment performed at GSI~\cite{gsi119a}, the cross section limit is below 50 fb at $E^{*}$$\sim$43 MeV.
Based on our calculations, the chances to discover 119 could be much larger with a lower beam energy at 3$n$ channel.
There are also predictions that the optimal reaction is at 4$n$ channel~\cite{Nasirov,zhang}.
For the $^{243}$Am+$^{54}$Cr reaction, the estimated cross section is about 11 fb at at $E^{*}$$\sim$37.5 MeV in the
3$n$-evaporation channel. This is an extremely low cross section for experiments. 
For comparison, another recent theoretical study estimated that the cross section is 25 fb at $E^{*}$$\sim$33 MeV 
for the 3$n$ channel~\cite{kayumov}. 
The calculated optimal beam energies $E_{\rm cm}$ are 226.8 MeV and 245.3 MeV for  $^{249}$Bk+$^{50}$Ti and  $^{243}$Am+$^{54}$Cr
reactions, respectively, 
based on the reaction $Q$ values given in our previous work~\cite{guan}. 
There are extensive studies of $^{249}$Bk+$^{50}$Ti and  $^{243}$Am+$^{54}$Cr cross sections, as summarized in ~\cite{zhang}.
The uncertainties in optimal beam energies in the same channel are generally less than 4 MeV. 
In addition, the fast fission could play a role to reduce the cross sections at high angular momentum~\cite{kayumov},
and in this respect the microscopic dynamical study of fusion process is needed. Note that the fast fission of a mononucleus is later than the quasifission,
as explained in~\cite{kayumov}. 
The machine learning methods~\cite{yuan,he}
might also be useful to infer the optimal collision energy for production of superheavy nuclei.

\emph{Summary.}---
The survival probabilities of compound superheavy nuclei are studied
based on microscopic energy dependent fission barriers, aiming
to synthesize the highly concerned new element Z=119. 
The triaxial deformation plays a significant role in reducing
the fission barriers of Z=119 isotopes  even at high excitation energies.
Together with the fusion probability from the DNS model,
the residual cross sections after the evaporation of multiple neutrons
are obtained.
This is a long-expected work since the microscopic studies of energy dependent fission barriers more than a decade ago. 
The cross sections of $^{243}$Am($^{48}$Ca, 3$n$)$^{288}$Mc can be reasonably described by adjusting $a_f/a_0$.
Our calculations indicate that the  $^{249}$Bk+$^{50}$Ti reaction with
a lower beam energy is promising to synthesize Z=119 element.


\acknowledgments
We are grateful to discussions with S.G Zhou and
 F.R. Xu, and discussions in a series of workshops on the synthesis of superheavy nuclei. 
 This work was supported by  the
 National Key R$\&$D Program of China (Grant No.2023YFA1606403, 2023YFE0101500),
  the National Natural Science Foundation of China under Grants No.12335007 and 11961141003.
We also acknowledge the funding support from the State Key Laboratory of Nuclear Physics and Technology, Peking University (No. NPT2023ZX01).


\begin{thebibliography}{99}

\bibitem{118}
Yu. Ts. Oganessian et al.,
Phys. Rev. C 74, 044602(2006).


\bibitem{Fe58}
Y. T. Oganessian  et al.,
Phys. Rev. C 79, 024603 (2009).

\bibitem{V51}
K. Chapman, Hunt for Element 119 to Begin, Chemistry World
(September 12, 2017), https://www.chemistryworld.com/news/
hunt-for-element-119-to-begin/3007977.article.

\bibitem{gsi119}
S. Hofmann, J. Phys. G 42, 114001 (2015).

\bibitem{gsi119a}
J. Khuyagbaatar, A. Yakushev, C. E. Dullmann, Ch. E.
Dullmann, D. Ackermann, L.-L. Andersson et al., Phys. Rev.
C 102, 064602 (2020).

\bibitem{gsi120}
 S. Hofmann, A. Heinz, R. Mann, J. Maurer, G. Munzenberg, S.
Antalic et al., Eur. Phys. J. A 52, 180 (2016).


\bibitem{Mc286}
Yu. Ts. Oganessian et al.,
Phys. Rev. C 106, 064306(2022).

\bibitem{cafe2}
W. Lu, H.Y. Ma, C. Qian, J.D. Ma, W.H. Zhang, Y.C. Feng, Z.H. Zhang, Z.H. Jia, L.B. Li, X. Fang, P. Zhang, H. Zhang, J.J. Chang, X.Z. Zhang, L.T. Sun, Y. He, H.W. Zhao, Nucl. Inst. Methods Phys. Res. A, 1062, 169207(2024).

\bibitem{pei2009}
J. C. Pei, W. Nazarewicz, J. A. Sheikh, and A. K. Kerman, Phys.
Rev. Lett. 102, 192501 (2009).

\bibitem{zhuy}
Y. Zhu and J. C. Pei, Phys. Rev. C 94, 024329 (2016).

\bibitem{itkis}
M. G. Itkis, Yu. Ts. Oganessian, and V. I. Zagrebaev,
Phys. Rev. C 65, 044602(2002).

\bibitem{qiaocy}
C. Y. Qiao and J. C. Pei,
Phys. Rev. C 106, 014608(2022).

\bibitem{moller}
P. Möller, A. J. Sierk, T. Ichikawa, A. Iwamoto, R. Bengtsson, H. Uhrenholt, and S. Åberg,
Phys. Rev. C 79, 064304(2009).

\bibitem{caiqz}
Qing-Zhen Chai, Wei-Juan Zhao, Min-Liang Liu, Hua-Lei Wang, Chinese Phys. C 42, 054101(2018).


\bibitem{lvbn}
Bing-Nan Lu, Jie Zhao, En-Guang Zhao, and Shan-Gui Zhou,
Phys. Rev. C 89, 014323 (2014).


\bibitem{geng}
Chang Geng, Peng-Hui Chen, Fei Niu, Zu-Xing Yang, Xiang-Hua Zeng, and Zhao-Qing Feng,
Phys. Rev. C 109, 054611 (2024) 


\bibitem{guan}
D.W. Guan, J.C. Pei, Phys. Lett. B 851,138578(2024).


\bibitem{deng}
Xiang-Quan Deng, Shan-Gui Zhou, Phys. Rev. C 107, 014616 (2023).


\bibitem{goodman}
A. L. Goodman, Nucl. Phys. A 352, 30 (1981).

\bibitem{hfodd}
N. Schunck, J. Dobaczewski, W. Satuła, P. Baczyk, J. Dudek, Y. Gao, M. Konieczka, K. Sato, Y. Shi, X.B. Wang, and T.R. Werner,
Comput.Phys. Comm. 216, 145(2017).

\bibitem{yshi}
Chen Ling, Chao Zhou, Yue Shi,  Eur. Phys. J. A 56, 180 (2020).

\bibitem{unedf1}
M. Kortelainen, J. McDonnell, W. Nazarewicz, P.-G. Reinhard, J. Sarich, N. Schunck, M. V. Stoitsov, and S. M. Wild,
Phys. Rev. C 85, 024304 (2012).


\bibitem{mix-pair}
{J. Dobaczewski, W. Nazarewicz, and M.V. Stoitsov, Nuclear ground-state properties from mean-field calculations,
Eur. Phys. J. A {\bf 15}, 21
  (2002)}.
  
\bibitem{bayesian}
{J.D. McDonnell, N. Schunck, D. Higdon, J. Sarich, S.M. Wild, and W. Nazarewicz, 
Phys. Rev. Lett. 114, 122501 (2015)  }

\bibitem{fengzq}
Zhao-Qing Feng, Gen-Ming Jin Fen Fu, Jun-Qing Li,  Nucl. Phys. A 771, 50(2006).

\bibitem{xia}
C. Xia, B. Sun, E. Zhao, and S. Zhou, Sci. China: Phys., Mech.
Astron. 54, 109 (2011).


\bibitem{zhang}
Ming-Hao Zhang, Yu-Hai Zhang, Ying Zou, Chen Wang, Long Zhu, and Feng-Shou Zhang,
Phys. Rev. C 109, 014622(2024).


\bibitem{zubov}
A. S. Zubov, G. G. Adamian, N. V. Antonenko, S. P. Ivanova, and W. Scheid,
Phys. Rev. C 65, 024308(2002).  
  
\bibitem{Ignatyuk}
A. V. Ignatyuk, M. G. Itkis, V. N. Okolovich, G. N. Smirenkin,
and A. S. Tishin, Yad. Fiz 21, 1185 (1975) [Sov. J. Nucl. Phys.
21, 612 (1975)].  
  

\bibitem{lijx}
Jia-Xing Li, and Hong-Fei Zhang,
Phys. Rev. C 108, 044604 (2023)

\bibitem{kayumov}
B. M. Kayumov, O. K. Ganiev, A. K. Nasirov, and G. A. Yuldasheva,
Phys. Rev. C 105, 014618 (2022)

\bibitem{fanli}
Fan Li, Long Zhu, Zhi-Han Wu, Xiao-Bin Yu, Jun Su, and Chen-Chen Guo,
Phys. Rev. C 98, 014618 (2018).

\bibitem{adamian}
G. G. Adamian, N. V. Antonenko, and W. Scheid
Phys. Rev. C 69, 044601 (2004).


\bibitem{Nasirov}
A. Nasirov and B. Kayumov,
Phys. Rev. C 109, 024613(2024).

\bibitem{denisov}
V. Yu. Denisov,
Phys. Rev. C 109, 044618(2024).  



\bibitem{sheikh}
J. A. Sheikh, W. Nazarewicz, and J. C. Pei
Phys. Rev. C 80, 011302(R) (2009).
  
\bibitem{mm-barrier}
M. Kowal, P. Jachimowicz, and A. Sobiczewski,
Phys. Rev. C 82, 014303(2010)

\bibitem{wangning}
Ning Wang and Min Liu, arXiv:2404.14014(2024).

\bibitem{skm}
J. Bartel, P. Quentin, M. Brack, C. Guet, and H. B. H{\aa}kansson, 
Nucl. Phys. A {\bf 386}, 79 (1982).


\bibitem{tub}
T. Bürvenich, M. Bender, J. A. Maruhn, and P.-G. Reinhard,
Phys. Rev. C 69, 014307(2004).

\bibitem{qiangy}
Yu Qiang and J. C. Pei,
Phys. Rev. C 104, 054604 (2021) 

\bibitem{bass}
R. Bass, Lecture Notes in
Physics Vol. 117 (1980), p. 281.

\bibitem{oganessian24}
Yu. Ts. Oganessian et al.,
Phys. Rev. C 109, 054307 (2024).


\bibitem{yuan}
B.S. Cai and C.X. Yuan, NUCL. SCI. TECH. 34, 204 (2023)


\bibitem{he}
Wanbing He, Qingfeng Li, Yugang Ma, Zhongming Niu, Junchen Pei, Yingxun Zhang,
Sci. China Phys. Mech. Astron. 66, 282001 (2023).
  

















\end{thebibliography}
\end{document}